\documentclass[french]{pfia}

\usepackage[labelfont=bf,textfont=it,labelsep=period,justification=raggedright,singlelinecheck=false]{caption}
\usepackage{xcolor}
\usepackage{hyperref}
\usepackage{comment}
\usepackage{listings}
\usepackage{algorithmic}
\usepackage{graphicx}
\usepackage{textcomp}
\usepackage{seqsplit}
\usepackage{tikz}
\usepackage{tabularx}
\usepackage{array}
\usepackage{adjustbox}
\usepackage{multirow}
\usepackage{pgfplots}
\usepackage{titlesec}
\usepackage{tablefootnote}
\usepackage{setspace}
\titlespacing{\section}{0pt}{0.5ex plus 0.5ex minus .2ex}{0.5ex}
\titlespacing{\subsection}{0pt}{0.5ex plus 0.5ex minus .2ex}{0.5ex}
\titlespacing{\subsubsection}{0pt}{0.5ex plus 0.4ex minus .2ex}{0.5ex}

\definecolor{afiablue}{RGB}{61,159,207}
\definecolor{afiared}{RGB}{167,75,68}
\definecolor{afialightblue}{RGB}{158,193,232}

\newcommand{\field}[1]{\textcolor{afiablue}{#1}}
\newcommand{\val}[1]{\textsf{\textcolor{afialightblue}{#1}}}
\renewcommand{\arraystretch}{1.5}
\usepackage{caption}
\title{\textbf{Vers un cadre ontologique pour la gestion des compétences : à des fins de formation, de recrutement, de métier, ou de recherches associées}}

\author{Ngoc Luyen Le\textsuperscript{1,2},  Marie-Hélène Abel\textsuperscript{2}, Bertrand Laforge\textsuperscript{1,3}\\[4pt]
	\footnotesize\textsuperscript{1}Gamaizer, 93340 Le Raincy, France \\[-8pt]
	\footnotesize\textsuperscript{2}Université de technologie de Compiègne, CNRS, Heudiasyc (Heuristics and Diagnosis of Complex Systems), CS 60319 - 60203 Compiègne Cedex, France\\[-8pt]
	\footnotesize\textsuperscript{3}Sorbonne Université, CNRS UMR 7585, LPMHE (Laboratoire de Physique Nucléaire et des Hautes Énergies), 75252 Paris cedex 05, France\\[-8pt]
}

\date{}

\begin{document}

\maketitle


\begin{resume}
La transformation rapide du marché du travail, alimentée par les avancées technologiques et l'économie numérique, exige un développement continu des compétences et une adaptation constante. Dans ce contexte, les systèmes traditionnels de gestion des compétences manquent d'interopérabilité, d'adaptabilité et de compréhension sémantique, rendant difficile l'alignement des compétences individuelles avec les besoins du marché du travail et les formations. Cet article propose un cadre basé sur l'ontologie pour la gestion des compétences, permettant une représentation structurée à différents niveaux de granularité, allant des micro-capacités aux macro-compétences, afin de répondre à des besoins liés aux applications métiers ainsi qu'à la structuration des compétences dans les contextes professionnels et de formation. En exploitant des modèles ontologiques et un raisonnement sémantique, ce cadre vise à améliorer l'automatisation de l'appariement compétences-métiers, la personnalisation des recommandations d'apprentissage et la planification de carrière. Cette étude discute de la conception, de la mise en œuvre et des applications potentielles du cadre, s'appuyant sur la recherche de formation des compétences, la recherche d'un emploi et la recherche de personnes compétentes.
\end{resume}

\begin{motscles}
Ontologies, Base de connaissances, Cadre ontologique, Gestion de compétences, Granularité de compétence.
\end{motscles}

\begin{abstract}
The rapid transformation of the labor market, driven by technological advancements and the digital economy, requires continuous competence development and constant adaptation. In this context, traditional competence management systems lack interoperability, adaptability, and semantic understanding, making it difficult to align individual competencies with labor market needs and training programs. This paper proposes an ontology-based framework for competence management, enabling a structured representation of competencies, occupations, and training programs. By leveraging ontological models and semantic reasoning, this framework aims to enhance the automation of competence-to-job matching, the personalization of learning recommendations, and career planning. This study discusses the design, implementation, and potential applications of the framework, focusing on competence training programs, job searching, and finding competent individuals.
\end{abstract}



\section{Introduction}
La transformation rapide du marché du travail, portée par les avancées technologiques et la transition vers une économie numérique, impose une évolution constante des compétences professionnelles ainsi qu’une annotation plus fine et performante des compléments au diplôme. Les approches traditionnelles de gestion des compétences rencontrent plusieurs défis, notamment le manque d'interopérabilité, la rigidité des taxonomies et l'absence d'une représentation sémantique permettant un alignement efficace entre les compétences des individus, celles liées aux offres d'emploi et celles associées aux formations disponibles~\cite{brittain2015competency,arribas2024systematic}. Cette inadéquation complique la mise en place de stratégies de développement des compétences adaptées aux besoins du marché du travail. Au-delà des enjeux professionnels, cette difficulté existe également dans l’accompagnement pédagogique des étudiants tout au long de leur formation. Il s’agit notamment de mieux relier les compétences ou capacités spécifiques qu’ils mobilisent dans une activité pédagogique à l’ensemble des macro-compétences visées par la formation.

Face à ces défis, l'adoption d'une approche ontologique apparaît comme une solution innovante permettant de modéliser et de relier, de manière structurée, les connaissances relatives aux compétences, aux métiers et aux formations~\cite{abel2007apport,miranda2017ontology}. En s'appuyant sur des standards reconnus (ROME\footnote{ROME : Répertoire Opérationnel des Métiers et des Emplois, \href{https://www.francetravail.org/opendata/repertoire-operationnel-des-meti.html?type=article}{https://www.francetravail.org} }, ESCO\footnote{ESCO: European Skills, Competences, Qualifications and Occupations, \href{https://ec.europa.eu/esco}{https://ec.europa.eu/esco}})
, il devient possible de développer un cadre pour la gestion des compétences. Ce cadre favorise l'interopérabilité entre différents systèmes (plateformes de formation, bases de données de métiers et d'offres d'emploi, etc.) et facilite la recommandation de formations, d'évolutions professionnelles ou de recrutement. En exploitant les capacités du web sémantique et des techniques de raisonnement automatisé, une ontologie permet de formaliser la représentation des compétences, des métiers et des formations~\cite{draganidis2006ontology,miranda2017ontology}. Cette approche vise à améliorer le rapprochement compétences-métiers, personnalise les recommandations de formation et anticipe mieux les évolutions des compétences requises.

Cet article présente un cadre ontologique pour la gestion des compétences, conçu pour répondre aux besoins actuels en matière de formation, de recrutement et d’évolution professionnelle. Ce cadre vise à assurer une meilleure correspondance entre les compétences des individus, les exigences du marché du travail, les opportunités d’emploi et les offres de formation. Basé sur une ontologie structurée, il modélise les compétences et établit leurs relations avec les métiers, tout en intégrant des mécanismes d’inférence sémantique et des algorithmes de recommandation. Ce dispositif permet d’identifier les formations les plus adaptées aux objectifs professionnels des utilisateurs, en facilitant leur montée en compétences et en renforçant leur employabilité.

Ce cadre s’inscrit dans les dynamiques actuelles de recherche sur la personnalisation des parcours éducatifs, l’orientation professionnelle assistée et la cartographie dynamique des compétences. Il fait notamment l’objet d’expérimentations dans des environnements tels que la plateforme de jeux éducatifs Ikigai.games\footnote{\url{https://ikigai.games/}}, qui propose des scénarios interactifs visant à renforcer l’engagement des apprenants et le développement de compétences transversales dans des contextes variés.

La suite de cet article est organisée comme suit : la section suivante propose une revue des travaux connexes dans le domaine de la gestion des compétences et des ontologies appliquées. Elle est suivie d'une description détaillée de la problématique et du cadre ontologique proposé, incluant sa structure conceptuelle et les aspects liés à son implémentation. Nous présentons alors une étude de cas basée sur le référentiel ROME 4.0 afin d'illustrer notre approche. Enfin, l'article se conclut par une synthèse des principaux résultats obtenus et une discussion sur les perspectives futures de recherche et d'application.
\renewcommand{\arraystretch}{1.0}
\begin{table}[h]
	\vspace{-0.3cm}
	\scriptsize
	\centering
	\begin{tabular}{|p{0.7cm}|p{1.8cm}|p{2.0cm}|p{2.1cm}|}
		
		\hline
		\textbf{Cadre} & \textbf{Couverture} & \textbf{Capacités Sémantiques} & \textbf{Cas d'Utilisation} \\ \hline
		ESCO Européen & Emplois, Compétences, Qualifications & RDF sans formalisation OWL & Planification de la main-d'œuvre, Éducation \\ \hline
		O*NET États-Unis & Professions et Compétences professionnelles & Taxonomie de base & Appariement emploi, Planification de carrière \\ \hline
		ROME France & Professions et Compétences professionnelles & RDF sans formalisation OWL & Services d'emploi \\ \hline
		\tiny{HR-XML Global} & Échange de données RH & Aucun support sémantique & Intégration des systèmes RH \\ \hline
		RNCP France & Certifications et qualifications professionnelles & Classification normalisée : référentiel reconnu officiellement & Validation des acquis, Formation professionnelle \\ \hline
		SFIA Global & Compétences numériques et informatiques & Modèle structuré des compétences & Gestion des compétences IT, Développement professionnel \\ \hline
		ISCO Global & Professions et compétences professionnelles & Classification hiérarchique : organisation des concepts en niveaux & Comparaison internationale, Statistiques sur l'emploi \\ \hline
	\end{tabular}
	\caption{Comparaison des Taxonomies et Standards de Compétences}
	\label{tab:comparaison_taxonomies}
	\vspace{-0.5cm}
\end{table}
\section{Travaux de la littérature}
La gestion des compétences a été largement explorée dans les domaines de l’éducation, des ressources humaines et de l’IA. Deux approches dominent : les taxonomies classiques et les modèles ontologiques.

Les taxonomies (ESCO, ROME, O*NET, etc.) permettent de classifier les compétences et métiers. La Table~\ref{tab:comparaison_taxonomies} résume leurs portées, formats et usages. Bien qu'utiles pour la standardisation, ces modèles restent statiques et peu exploitables par des systèmes intelligents.

Pour répondre à ces limites, plusieurs travaux ont proposé des ontologies afin d’offrir une structuration dynamique, formelle et interopérable des compétences. Ces modèles permettent des inférences et recommandations automatisées. Par exemple, Miranda et al.~\cite{miranda2017ontology} ont conçu une ontologie pour les systèmes RH, tandis que Paquette~\cite{paquette2007ontology,paquette2021new} a modélisé les compétences en contexte éducatif. D'autres travaux~\cite{draganidis2006ontology,rezgui2014ontology} s'intéressent à la personnalisation des parcours.

Cependant, peu de modèles relient explicitement compétences, métiers et formations, ou exploitent pleinement les règles d'inférence pour la recommandation. Notre travail vise à répondre à ces lacunes en proposant un cadre unifié, interopérable et orienté vers les cas d’usage formation/recrutement.

\section{Problématique}
Dans un contexte où la gestion des compétences doit s’adapter en permanence aux évolutions du marché du travail, plusieurs défis émergent. Pour un individu souhaitant exercer un métier, il est essentiel d’identifier les compétences requises et d’évaluer celles déjà acquises. Lorsqu’un écart de compétences est constaté, la question centrale est de déterminer quelles formations permettront d’acquérir les savoirs et savoir-faire nécessaires. Du point de vue des ressources humaines, le recrutement repose sur la capacité à évaluer les candidats en fonction des exigences d’un poste. Il s’agit de vérifier que la personne possède les compétences attendues ou, si nécessaire, de recommander des formations adaptées pour combler ces lacunes.

Un cadre ontologique structuré permettrait de modéliser et d’interconnecter les relations entre métiers, compétences et formations. Une telle approche faciliterait l’orientation professionnelle des individus, l’adéquation entre l’offre et la demande en compétences et l’optimisation des processus de recrutement. En exploitant les capacités des ontologies et du raisonnement sémantique, ce cadre contribuerait à une gestion plus efficace et dynamique des compétences, en anticipant les évolutions du marché, en améliorant la mobilité professionnelle et en renforçant l’adéquation entre les parcours éducatifs et les besoins du marché du travail. Dans la section suivante, nous présentons notre approche pour développer un cadre basé sur l'ontologie pour la gestion de compétences dans la prochaine section.

\section{Cadre de gestion des compétences basé sur l'ontologie}

Cette section décrit un cadre ontologique pour structurer les connaissances liées aux compétences, métiers et formations. Elle présente d'abord l’architecture en couches du système, puis détaille la modélisation ontologique des entités et leur intégration dans l’écosystème de gestion des compétences.

\subsection{Achitecture du cadre ontologique}
\begin{figure}[h!]
	\vspace{-0.2cm}
	\centering\includegraphics[width=0.3\textwidth]{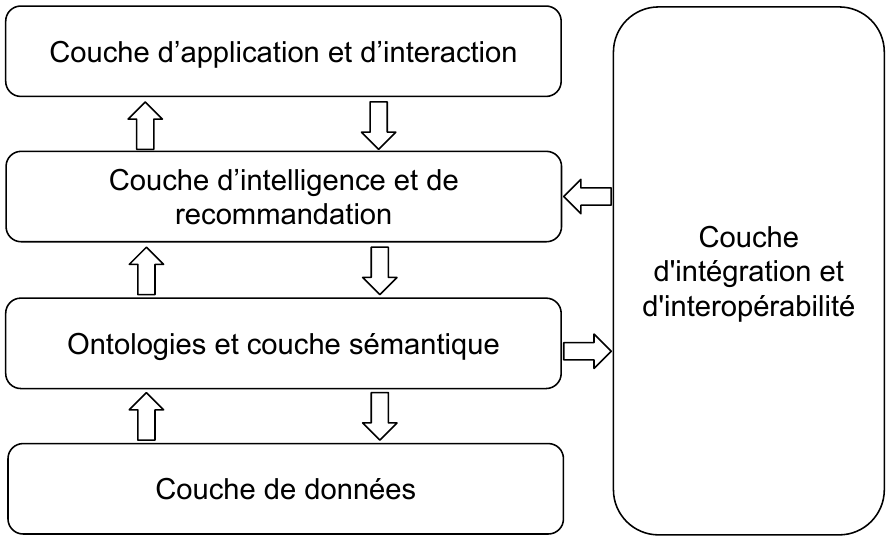}
	\vspace{-0.2cm}
	\caption{Architecture du Cadre Ontologique pour la gestion de compétence} \label{fig_01}
		\vspace{-0.3cm}
\end{figure}

Le cadre repose sur une architecture en couches assurant la structuration des données, le raisonnement sémantique, les recommandations intelligentes, et l’interopérabilité avec les systèmes externes (Fig.~\ref{fig_01}).

\textit{Couche de données} : regroupe compétences, métiers, formations et profils d'apprenants. Elle centralise les informations issues des référentiels et données contextuelles (comportementales, physiologiques) afin de structurer les ressources et personnaliser les parcours~\cite{sulema2023augmented}.

\textit{Couche de sémantique} : modélisation des entités (compétences, métiers, formations) en OWL/RDF, construction d’un graphe de connaissances aligné sur les référentiels. Inférences logiques (écarts, correspondances) utilisées par la couche suivante pour la recommandation~\cite{le2023improving}.

\textit{Couche d'intelligence et de recommandation} : moteur de recommandation basé sur les écarts de compétences, appariement profil-métier via IA, prédiction de trajectoires et suivi dynamique des compétences~\cite{dawson2021skill,le2023constraint,decorte2023career}.

\textit{Couche d'application et d'interaction} : interfaces pour apprenants (visualisation des compétences, recommandations), recruteurs (recherche de profils), et formateurs (gestion de l’offre pédagogique).

\textit{Couche d'intégration et d'interopérabilité} : connexion aux plateformes LMS (Moodle, Coursera), outils de gestion de resources(Memorae), jeux éducatif (Ikigai.games), et systèmes d’information d’entreprise. Données personnelles conformes RGPD (pseudonymisation, consentement, traçabilité).

\subsection{Ontologie de gestion des compétences}
Le développement d'une ontologie pour la gestion des compétences permet de structurer et de formaliser les relations entre les compétences, les formations et les métiers. Cette section présente l'ontologie CMO - \textbf{C}ompetency \textbf{M}anagement \textbf{O}ntology, conçue pour servir de support à la gestion des compétences et à leur intégration dans un écosystème global de gestion des talents.

\begin{figure}[h!]
	\vspace{-0.3cm}
	\centering\includegraphics[width=0.48\textwidth]{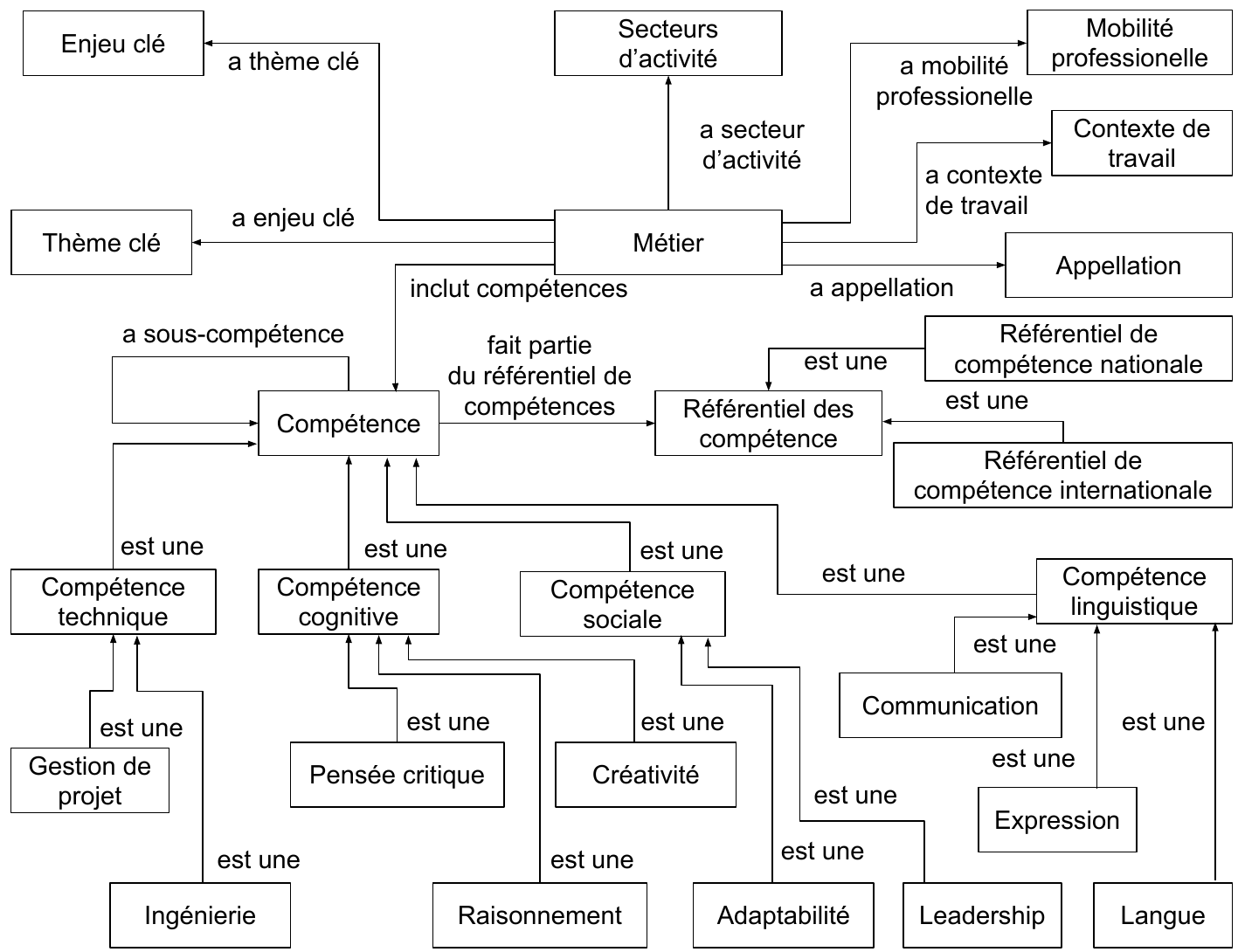}
	\caption{Modélisation ontologique des compétences et de leur contexte d'application} \label{fig_02}
	\vspace{-0.3cm}
\end{figure}

L'ontologie CMO repose principalement sur une structuration détaillée des compétences, prenant en compte leur classification ainsi que leur contexte métier. Comme illustré dans la figure \ref{fig_02}, les \textit{compétences} sont organisées en différentes catégories : compétences sociales (ex. Leadership, Communication), compétences cognitives (ex. Pensée critique, Raisonnement), compétences techniques (ex. Gestion de projet, Ingénierie) et compétences linguistiques (ex. Langues, Expression)~\cite{abel2007apport,foveau2007referentiels}. Chaque compétence peut être décomposée en sous-compétences, permettant ainsi une granularité plus fine dans l’évaluation et la structuration des connaissances et savoir-faire.
En complément, l’ontologie intègre les \textit{référentiels de compétences}, qui servent à normaliser la description et l’évaluation des compétences selon des standards établis. Elle distingue notamment les référentiels de compétence nationaux, définis par des institutions gouvernementales ou académiques, et les référentiels de compétence internationaux, alignés sur des modèles tels qu’ESCO, ROME, O*NET ou HR-XML. L’association des compétences à ces référentiels permet une meilleure compatibilité avec les cadres normatifs existants et facilite la reconnaissance des compétences au niveau international.

Les compétences sont directement liées aux métiers, permettant d’identifier les aptitudes nécessaires pour exercer une profession. Chaque métier est caractérisé par son secteur d’activité, qui définit son domaine économique, ainsi que par son contexte de travail, précisant les conditions et les exigences spécifiques associées à l’emploi. L’ontologie prend également en compte la mobilité professionnelle, en modélisant les transitions possibles entre métiers en fonction des compétences transférables. De plus, elle intègre des notions d’enjeux et de thèmes clés, mettant en lumière les tendances et évolutions qui influencent l’évolution des compétences dans un domaine donné.

Sur le plan structurel, l’ontologie repose sur des relations sémantiques qui définissent les interconnexions entre les entités. Parmi celles-ci, la relation ``\textit{a sous-compétence}' permet de détailler une compétence en sous-éléments et facilitant ainsi la représentation de compétences composites ou hiérarchisées, tandis que ``\textit{fait partie du référentiel de compétences}'' associe une compétence à un référentiel de compétences spécifique. La relation ``\textit{a secteur d'activité}'' établit le lien entre un métier et son domaine professionnel, et ``\textit{a enjeu clé}'' et ``\textit{a thème clé}'' identifient les défis et évolutions qui impactent les besoins en compétences d'un métier. Enfin, ``\textit{a contexte travail}'' décrit l’environnement dans lequel une compétence est mise en pratique.

\begin{figure}[h!]
	\centering\includegraphics[width=0.48\textwidth]{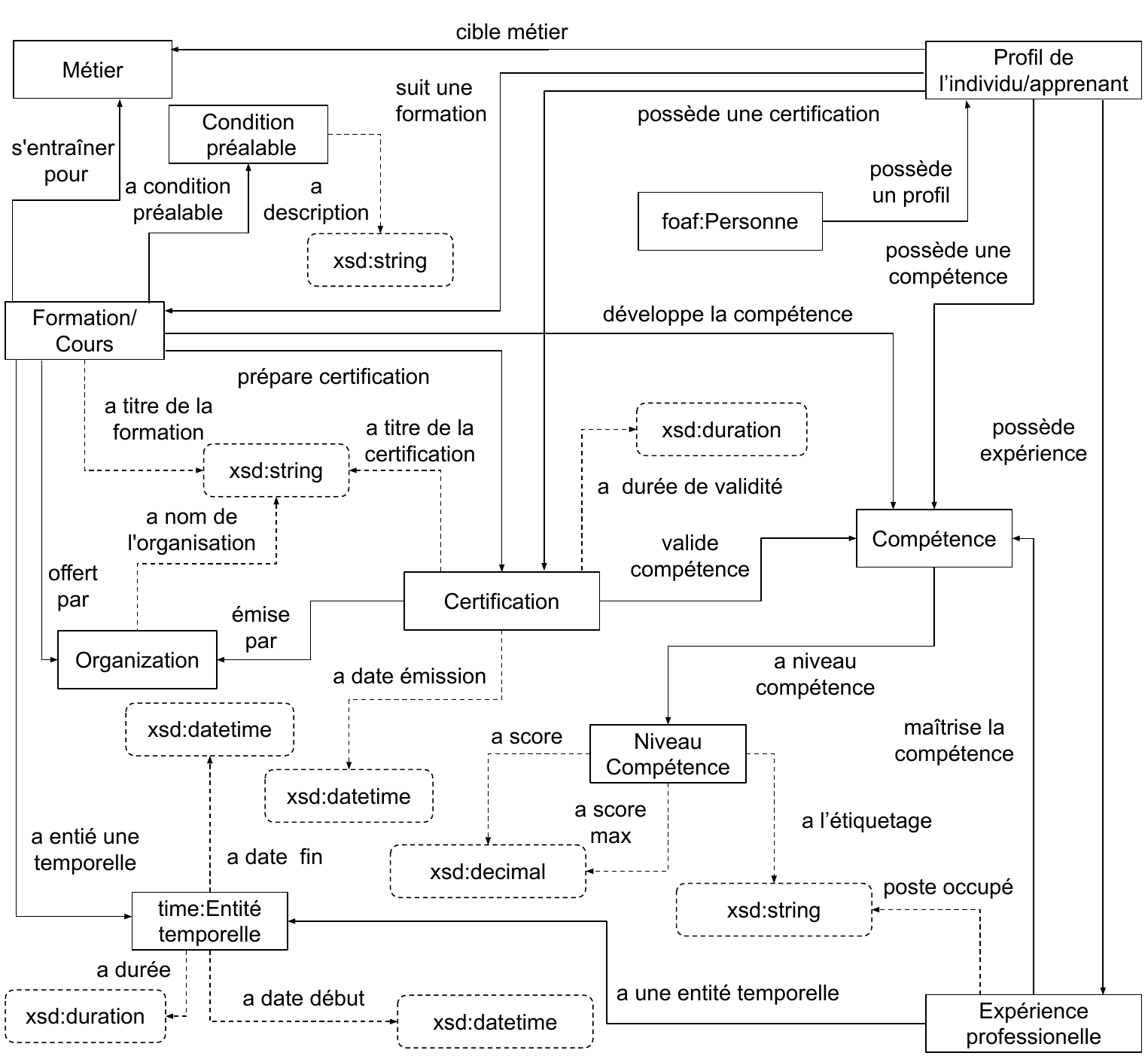}

	\caption{ Intégration des compétences avec les formations et les certifications} \label{fig_03}
		\vspace{-0.7cm}
\end{figure}

\`A cette structuration des compétences vient se positionner les profils des individus/apprenants, les formations ou cours suivis, et les certifications obtenues, et les expériences professionelles. Les entités temporelles et des niveaux de compétences viennent alors compléter le modèle de façon à permettre d'évaluer dynamiquement les profils, comme illustré dans la figure \ref{fig_03}. En particuler, chaque individu/apprenant possède un profil d'individu/apprenant, qui regroupe l’ensemble de ses compétences acquises soit par formation, soit par expérience professionnelle. Chaque compétence est associée à un niveau de compétence, permettant de quantifier l’expertise acquise par un individu/apprenant. Ce niveau peut être précisé à l’aide de scores et d’étiquetages sémantiques (``\textit{a niveau compétence}'', ``\textit{a score max}'', ``\textit{a score}''), offrant ainsi une granularité fine dans l’évaluation des compétences. 

Les formations jouent un rôle clé dans le développement des compétences. Un individu/apprenant peut suivre une formation dispensée par une organisation, celle-ci pouvant être une université, un centre de formation ou une entreprise. L’ontologie permet également de modéliser les conditions préalables (``\textit{a condition préalable}''), garantissant que l’individu dispose des préconditions minimales requises avant d’accéder à une formation donnée. En complément des formations, les certifications viennent valider officiellement certaines compétences. Une certification est émise par une organisation et est directement liée à une compétence validée (``\textit{valide compétence}''). Chaque certification possède une date d’émission (``\textit{a date émission}'') et une durée de validité (``\textit{a durée de validité}''), assurant ainsi une gestion efficace des qualifications professionnelles.

Un élément central de cette ontologie est le processus de validation des compétences. Lorsqu'un individu/apprenant complète une formation, il peut obtenir une certification reconnue, garantissant ainsi la légitimité de ses acquis auprès des employeurs et des institutions. Cette certification est délivrée par une organisation accréditée et permet de formaliser la montée en compétence de l'individu. L’acquisition des compétences ne se limite pas aux formations ; elle peut également être consolidée à travers l’expérience professionnelle. Un individu possède une expérience professionnelle, qui est associée à un \textit{poste occupé}. Cette expérience permet de maîtriser des compétences spécifiques et constitue un facteur clé dans l’évaluation des aptitudes professionnelles.

L’intégration d’une entité temporelle est un aspect fondamental de cette ontologie. Chaque formation, certification et expérience professionnelle est associée à une durée (``\textit{xsd:duration}''), une date de début (``\textit{a date début}'') et une date de fin (``\textit{a date fin}''). Cette structuration temporelle permet de suivre l’évolution des compétences d’un individu sur le long terme et d’optimiser la gestion des carrières. Afin de s’adapter à l’évolution rapide des métiers et des compétences, l’ontologie prévoit un processus de maintenance continue, reposant sur l’alignement régulier avec des référentiels actualisés (tels que ROME ou ESCO) et sur une validation collaborative par des experts métiers. Une interface d’édition permet en outre de proposer, réviser et valider de nouvelles compétences au fil du temps.
Pour illustrer concrètement l’application de l’ontologie CMO et du cadre associé, une étude de cas détaillée est présentée dans la section suivante.

\section{Étude de cas}
Dans cette section, nous nous concentrons sur l'expérimentation du cadre ontologique et de l'ontologie CMO à travers une étude de cas visant à requalifier des individus/apprenants vers des métiers en forte demande selon le référentiel compétences ROME 4.0. Nous présentons le contexte et les objectifs, détaillons le scénario de l'étude de cas, explorons l'interrogation ontologique à l’aide de requêtes SPARQL, et discutons des résultats attendus et de leur impact.

\subsection{Contexte et objectifs}
Dans un marché du travail en constante évolution, accéléré par l'intelligence artificielle générative~\cite{marty2024intelligence}, de nombreux individus/apprenants cherchent à améliorer leurs compétences pour accéder à des métiers en forte demande. Ces évolutions sont particulièrement notables dans le cadre du ROME 4.0, qui propose une classification des métiers en fonction des compétences requises, permettant ainsi une meilleure correspondance entre l’offre et la demande.
Dans ce contexte, nous proposons une approche ontologique permettant de modéliser les compétences, les parcours de formation et les exigences des métiers. Ce cadre formel facilite l’identification des écarts entre les compétences détenues par un individu et celles exigées par un métier cible. Grâce à cette approche, il devient possible de recommander des formations personnalisées, d’intégrer des systèmes de certification et d’assurer une interopérabilité avec des plateformes de gestion des talents et d’apprentissage en ligne.


\subsection{Scénario de l'étude de cas}
Dans ce scénario, nous considérons \textit{Louis Le}, un apprenant souhaitant évoluer vers le métier de \textit{Data Scientist} (Code du \textit{ROME M1405}\footnote{\url{https://candidat.francetravail.fr/metierscope/fiche-metier/M1405/data-scientist}}). Actuellement, il possède des compétences en Python, mais uniquement à un niveau basique (\textit{cmo:Niveau01}). Cependant, le métier de \textit{Data Scientist} requiert des compétences avancées, notamment en \textit{Python Avancé}, en \textit{Machine Learning} et en \textit{Analyse de Données}, que \textit{Louis Le} ne maîtrise pas encore.

\begin{figure}[h!]
	\vspace{-0.3cm}
	\centering\includegraphics[width=0.48\textwidth]{ontology\_instance.pdf}
	\caption{Instance ontologique illustrant la relations d'un individu/apprenant vers un métier du ROME 4.0 en reliant ses compétences, et la formation recommandée.} \label{fig_04}
	
	\vspace{-0.3cm}
\end{figure}

L’ontologie permet d’analyser son profil actuel et de le comparer aux exigences du métier ciblé, identifiant ainsi un écart de compétences. Afin de combler cet écart, une formation adaptée est recommandée : la ``\textit{Formation Data Scientist DS25}, proposée par \textit{OpenClassrooms}. Comme illustré dans la Figure~\ref{fig_04}, cette formation est spécifiquement conçue pour permettre à \textit{Louis Le} d’acquérir les compétences manquantes.

Une fois inscrit à la ``\textit{Formation Data Scientist DS25}'', \textit{Louis Le} pourra progressivement améliorer son niveau en \textit{Python}, développer son expertise en \textit{Analyse de Données} et acquérir de nouvelles compétences en \textit{Machine Learning}. La formation, d’une durée de \textit{6 mois} et commencée le \textit{10/01/2025}, est structurée en plusieurs modules et est associée à un système d’évaluation des compétences.

À l'issue du programme, \textit{Louis Le} devra valider ses acquis grâce à une évaluation formelle (\textit{cmo:Niveau02} pour un niveau avancé en Python, par exemple). Si l'évaluation est réussie, une \textit{certification} lui sera délivrée, attestant de son niveau en Data Science et facilitant son insertion professionnelle.
L’intégration de cette ontologie avec des plateformes de certification et des systèmes RH permettra d’automatiser la mise à jour de son profil professionnel, garantissant ainsi une meilleure employabilité et une meilleure adéquation avec les offres du marché.

\subsection{Requêtes SPARQL sur l’ontologie}
L’utilisation des requêtes SPARQL permet d’extraire des informations clés sur les compétences d’un individu, les formations disponibles et les écarts à combler. Afin d’illustrer cette approche, nous proposons des requêtes visant à récupérer l’ensemble des compétences possédées par un individu, ainsi que leur niveau associé, et à identifier les compétences manquantes par rapport aux exigences du métier visé.

Dans la première requête SPARQL, illustrée dans la boîte de requête~\ref{lst:competence_extraction}, l’objectif est de récupérer l’ensemble des compétences possédées par un individu, ainsi que leur niveau associé, si cette information est disponible.

Plus précisément, la requête interroge l’ontologie en exploitant le prédicat \textit{possède une compétence -- cmo:possedeCompetence}, qui établit une relation entre un individu (\textit{Profil de l'apprenant -- cmo:ProfilApprenant}) et les compétences qu’il a acquises. Par ailleurs, une jointure optionnelle est effectuée à l’aide de la clause OPTIONAL, permettant de récupérer, lorsque disponible, le niveau de maîtrise correspondant à chaque compétence (\textit{a niveau compétence -- cmo:aNiveauCompetence}). Cette approche garantit que les individus dépourvus d’une information explicite sur leur niveau de compétence ne soient pas exclus des résultats.

\begin{lstlisting}[basicstyle=\scriptsize\ttfamily, language=SPARQL, caption=Requête SPARQL pour récupérer les compétences et niveaux associés d’un individu, breaklines=true, label=lst:competence_extraction] 
PREFIX cmo:  <http://gamaizer.ia/cmo#>
PREFIX rdf: <http://w3c.org/1999/02/22-rdf-syntax-ns#>
SELECT ?pa ?competence ?niveau 
WHERE { 
 ?pa rdf:type cmo:ProfilApprenant; 
 cmo:possedeCompetence ?competence .
 OPTIONAL {
  ?competence cmo:aNiveauCompetence
    ?niveau . 
 }
}
\end{lstlisting}
\vspace{-0.3cm}


\begin{table}[h!]
	\scriptsize
	\centering
	\caption{Résultats de la requête SPARQL : Extraction des compétences et niveaux pour différents individus}
	\label{tab:competences_individus}
	\begin{tabular}{|c|c|c|}
		\hline
		\textbf{Individu} & \textbf{Compétence} & \textbf{Niveau} \\ 
		\hline
		Louis Le  & Python 01               & Basique       \\ 
		\hline
		Henri Le & Python 01               & Avancé        \\ 
		Henri Le  & Machine Learning 02     & Intermédiaire \\ 
		Henri Le  & Big Data             & \textit{Non défini} \\ 
		\hline
		Marc   & Cybersécurité        & Expert        \\ 
		Marc   & Réseau Informatique  & Intermédiaire \\ 
		Marc   & Python 01               & Débutant      \\ 
		\hline
		Sophie & UX/UI Design         & Avancé        \\ 
		Sophie & Développement Web    & Intermédiaire \\ 
		Sophie & Python               & \textit{Non défini} \\ 
		\hline
	\end{tabular}
		\vspace{-0.6cm}
\end{table}

La Table~\ref{tab:competences_individus} présente les résultats de le requête SPARQL~\ref{lst:competence_extraction} extrayant les compétences et niveaux associés à différents individus. Chaque ligne associe un individu à une compétence et à son niveau de maîtrise, lorsqu’il est renseigné. Les résultats mettent en évidence des compétences sans niveau défini (\textit{Non défini}), indiquant des acquisitions non évaluées ou en attente de validation. Cette information est essentielle pour identifier les écarts de compétences et recommander des formations adaptées. Par exemple, \textit{Sophie}, experte en \textit{UX/UI Design}, pourrait renforcer son profil en complétant sa maîtrise de \textit{Python}, tandis que \textit{Marc}, débutant en \textit{Python}, pourrait bénéficier d’une montée en compétences ciblée.

Nous examinons une deuxième requête SPARQL, présentée dans la boîte de requête~\ref{lst:competence_missing}, dont l’objectif est d’identifier les compétences manquantes d’un individu souhaitant accéder à un métier cible. Plus précisément, cette requête compare les compétences requises pour le métier \textit{M1405} (\textit{Data Scientist}) avec celles déjà acquises par \textit{Louis Le}, en retournant uniquement celles qu’il ne possède pas encore.

\vspace{-0.2cm}
\begin{lstlisting}[basicstyle=\scriptsize\ttfamily, language=SPARQL, caption=Requête SPARQL pour identifier les compétences manquantes pour un métier cible,breaklines=true, label=lst:competence_missing]
PREFIX cmo: <http://gamaizer.ia/cmo#>
SELECT ?competenceRequise 
WHERE {
 cmo:M1405 cmo:includeCompetence ?competenceRequise .
 FILTER NOT EXISTS {
  cmo:LouisLe cmo:possedeUneCompetence ?competenceRequise .
 }
}
\end{lstlisting}
\vspace{-0.3cm}

La requête exploite la propriété \textit{include compétence -- cmo:includeCompetence}, qui associe un métier aux compétences nécessaires à son exercice. La clause FILTER NOT EXISTS permet d’exclure les compétences déjà détenues par \textit{Louis Le}, identifiées via la propriété \textit{possède une compétence -- cmo:possedeUneCompetence}. Ainsi, seules les compétences exigées par le métier et absentes du profil de l’apprenant seront affichées dans les résultats.

\vspace{-0.2cm}
\begin{table}[h!]
	\scriptsize
	\centering
	\caption{Résultats de la requête SPARQL : Identification des compétences manquantes pour le métier M1405 }

	\label{tab:competences_manquantes}
	\begin{tabular}{|c|c|}
		\hline
		\textbf{Individu} & \textbf{Compétence Manquante} \\ 
		\hline
		Louis Le  & Machine Learning 01       \\ 
		Louis Le  & Analyse de données 01    \\ 
		Louis Le  & Python\_02             \\ 
		\hline
	\end{tabular}
	\vspace{-0.3cm}
\end{table}

La Table~\ref{tab:competences_manquantes} présente les résultats de la requête SPARQL visant à identifier les compétences manquantes de \textit{Louis Le} pour accéder au métier \textit{M1405} (\textit{Data Scientist}). L’analyse des écarts de compétences révèle que \textit{Louis Le} ne possède pas encore les compétences clés suivantes : \textit{Machine Learning 01}, \textit{Analyse de données 01} et \textit{Python 02}.

Ces résultats indiquent que, bien que \textit{Louis Le} ait déjà certaines compétences en programmation, il doit encore acquérir des connaissances avancées en Python (\textit{Python 02}), ainsi qu’une maîtrise des techniques d’\textit{analyse de données} et de \textit{machine learning}. Ces compétences étant essentielles pour le poste visé, leur absence constitue un frein à son évolution vers le métier de \textit{Data Scientist}.

Grâce à cette détection automatique, il est possible d’orienter \textit{Louis Le} vers des formations spécifiques qui combleront ses lacunes. Par exemple, une formation avancée en \textit{Python} pour la \textit{Data Science}, un cours en \textit{Machine Learning} et une formation en \textit{Analyse de Données} seraient des recommandations pertinentes pour renforcer son profil. Cette approche permet d’éviter des apprentissages redondants, de raccourcir le parcours de requalification et d’accélérer son intégration dans le marché du travail.

Les deux requêtes SPARQL présentées permettent d’analyser le profil de compétences d’un individu en identifiant à la fois les compétences acquises et celles manquantes pour un métier cible. La première requête extrait les compétences possédées par un apprenant ainsi que leur niveau de maîtrise, offrant ainsi un état des lieux précis de ses acquis. La seconde requête, quant à elle, compare ces compétences aux exigences du métier \textit{M1405} (\textit{Data Scientist}) et détecte automatiquement les écarts de compétences que l’apprenant doit encore acquérir.

\subsection{Résultats attendus et impact}

L'intégration de l'ontologie CMO vise à améliorer la personnalisation des parcours d'apprentissage et de requalification. En identifiant automatiquement les écarts entre compétences détenues et compétences requises, elle permet de recommander des formations ciblées, réduisant les redondances et optimisant le temps d’apprentissage.

Pour les apprenants, cela se traduit par un accompagnement individualisé et une meilleure employabilité. Pour les recruteurs, le système facilite l’identification de profils pertinents via une visualisation structurée des compétences, enrichie par l’analyse de niveaux, scores et expériences. Les fournisseurs de formation peuvent, de leur côté, adapter leur offre en fonction des besoins réels du marché.

L’interopérabilité avec des plateformes comme Moodle, LinkedIn, Memorae ou Ikigai.games garantit une synchronisation fluide des données, assurant une mise à jour continue des profils. Le système respecte par ailleurs les exigences du RGPD grâce à des mécanismes intégrés de consentement, pseudonymisation et traçabilité.

Enfin, le cadre permet un suivi dynamique des compétences : les recommandations évoluent avec le marché grâce à l’analyse sémantique et aux contributions d’experts métiers. Il constitue ainsi un outil stratégique pour la gestion des talents, la mobilité professionnelle et l’anticipation des besoins futurs en compétences.

\section{Conclusion et perspectives}
Dans cet article, nous avons conçu un cadre ontologique pour la gestion des compétences, à des fins de formation, de recruitement, ou de métier, afin d’optimiser l’alignement entre les exigences du marché du travail et les formations disponibles. En intégrant des mécanismes de recherche et d'inférence basés sur l'ontologie, cette approche permet d’identifier les écarts de compétences et de recommander des formations adaptées, améliorant ainsi l’employabilité et accélérant la requalification professionnelle.
Ce cadre a fait l’objet d’une première validation exploratoire, posant les bases du développement d’un prototype et démontrant son potentiel pour structurer et automatiser la gestion des compétences. Plusieurs axes d’amélioration restent à explorer. L’intégration de modèles d’apprentissage automatique pourrait affiner les recommandations, en analysant plus finement l’évolution des compétences et les tendances du marché. Par ailleurs, une validation à plus grande échelle, associée à une meilleure interopérabilité avec les plateformes de formation et de recrutement, permettrait d’élargir l’impact du cadre et de renforcer son adoption. Des expérimentations futures sont ainsi envisagées pour évaluer sa robustesse, sa précision en contexte réel, ainsi que son acceptabilité par les utilisateurs.
En combinant ontologies, inférence sémantique et intelligence artificielle, ce cadre évolutif soutient l’adaptation continue des compétences et facilite les transitions professionnelles dans un marché en mutation.

\section*{Remerciements}
Nous remercions chaleureusement le consortium Ikigai porté par l’association Games for Citizens, la société Gamaizer ainsi que le projet FORTEIM (projet lauréat AMI CMA France 2030), pour leur soutien et leur collaboration. Leurs contributions ont apporté une valeur ajoutée significative à la réalisation de cette recherche.

\footnotesize
\bibliographystyle{plain}
\bibliography{references}

\end{document}